\documentclass[12pt,preprint]{aastex}

\def\alwaysmath#1{\ifmmode{#1}\else{$#1$}\fi} 

\def\ocen{\alwaysmath{\omega}~Centauri}

\def\hst{{\it HST}}
\def\HST{\hst}
\def\ltsima{$\; \buildrel < \over \sim \;$}
\def\gtsima{$\; \buildrel > \over \sim \;$}
\def\lsim{\lower.5ex\hbox{\ltsima}}
\def\gsim{\lower.5ex\hbox{\gtsima}}
\def\lapp{\ifmmode\stackrel{<}{_{\sim}}\else$\stackrel{<}{_{\sim}}$\fi}
\def\gapp{\ifmmode\stackrel{>}{_{\sim}}\else$\stackrel{<}{_{\sim}}$\fi}

\newdimen\minuswidth    %define @ width of minus sign for tables
\setbox0=\hbox{$-$}
\minuswidth=\wd0
\newdimen\digitwidth    %define ! a one digit width for tables
\setbox0=\hbox{\rm0}
\digitwidth=\wd0
\catcode`!=\active
\def!{\kern\digitwidth}
 
\shorttitle{BSS in $\omega$  Centauri} 
\shortauthors{Ferraro et al.} 

\begin{document}

\title{The pure non-collisional Blue Straggler population in the 
giant stellar system $\omega$ Centauri
\footnote{Based on observations with the NASA/ESA HST,
obtained at the Space Telescope Science Institute, which is operated
by AURA, Inc., under NASA contract NAS5-26555. Also based on WFI
observations collected at the European Southern Observatory,
La Silla, Chile, within the observing
programs 62.L-0354 and 64.L-0439.}
}
 
\author{F.R. Ferraro\altaffilmark{2}, A.
Sollima\altaffilmark{2}}
\affil{\altaffilmark{2}Dipartimento di Astronomia, Universit\`a di
Bologna, via Ranzani 1,I--40126 Bologna, Italy}
\email{francesco.ferraro3@unibo.it,antonio.sollima@bo.astro.it}

\author{R.~T.~Rood\altaffilmark{3}} 
\affil{\altaffilmark{3}Astronomy Dept., University of Virginia
Charlottesville VA~22903--0818, USA} 
\email{rtr@virginia.edu}
 
\author{ L. Origlia\altaffilmark{4}, E. Pancino
\altaffilmark{4}, M. Bellazzini\altaffilmark{4}} 
\affil{\altaffilmark{4} INAF-Osservatorio Astronomico 
 di Bologna, via Ranzani 1,I--40126 Bologna, Italy}
 \email{livia.origlia@bo.astro.it,elena.pancino@bo.astro.it
 michele.bellazzini@bo.astro.it}

\begin{abstract}

We have used high spatial resolution data from the Hubble Space
Telescope (\hst) and wide-field ground-based observations to search
for blue straggler stars (BSS) over the entire radial extent of the
large stellar system $\omega$ Centauri.  We have detected the largest
population of BSS ever observed in any stellar system. Even though the
sample is restricted to the brightest portion of the BSS sequence,
more than 300 candidates have been identified. BSS are thought to be
produced by the evolution of binary systems (either formed by stellar
collisions or mass exchange in binary stars). Since systems like
Galactic globular clusters (GGC) and \ocen\ evolve dynamically on
time-scales significantly shorter than their ages, binaries should have
settled toward the center, showing a more concentrated radial
distribution than the ordinary, less massive single stars. Indeed, in
all GGCs which have been surveyed for BSS, the BSS distribution is
peaked at the center. Conversely, in \ocen\ we find that the BSS share
the same radial distribution as the adopted reference populations.
This is the cleanest evidence ever found that such a stellar system is
not fully relaxed even in the central region.  We further argue that
the absence of central concentration in the BSS distribution rules out
a collisional origin. Thus, the \ocen\ BSS are the purest and largest
population of non-collisional BSS ever observed.  Our results allow
the first empirical quantitative estimate of the production rate of
BSS via this channel.  BSS in \ocen\ may represent the best local
template for modeling the BSS populations in distant galaxies where
they cannot be individually observed.  
\end{abstract}


\section{Introduction} 
\label{intro}

Blue straggler stars (BSS) define a sparsely populated sequence
extending to higher luminosity than the turnoff (TO) point of normal
hydrogen-burning main sequence (MS) stars in the color-magnitude diagrams
(CMD) of stellar aggregates like Galactic globular clusters (GGC).
Superficially they appear to be MS stars with masses larger than
expected for the cluster age as determined from the MSTO. There
are two  mechanisms thought to produce BSS: the first is mass
exchange in a binary system, the second is  merger of two stars
induced by stellar interactions, i.e., collisions (either between
single or binary stars) in a dense stellar environment. In either
scenario,  BSS are significantly more massive than  normal
cluster stars. Because of the high stellar density, 
collisions are more frequent in cluster centers. In
addition, clusters evolve dynamically, and more massive objects sink
toward the center on a time scale known as the relaxation time. Both
factors suggest that BSS should be more concentrated in the central
regions with respect to the other cluster stars, and this has been
found to be the case in all GGCs with adequate observations of the
center.
  
Among the stellar systems that populate the Galactic Halo, \ocen\ is
beyond any doubt the most surprising. The entire body of evidence
collected so far --- kinematics, spatial distribution and chemical
composition peculiarities --- make \ocen\ a unique object if classed as
usual among the globular clusters of the Milky Way (GGCs). It is more
massive ($M \approx 2.9 \times 10^6 M_{\sun}$, Merritt, Meylan \&
Mayor 1997) and luminous than any other GGC. It differs dynamically
from ordinary GGCs: it is one of the most flattened cluster
\citep{M87,WS87} and is partially supported by rotation
\citep{M87,MMM97}.  The most astonishing peculiarity of \ocen\ is its
metallicity spread measured both spectroscopically \citep{NFM96,SK96}
and photometrically \citep{L99,P00,HR00,Fr02,S05}. Being the only halo
stellar system with such a large chemical inhomogeneity, it has been
suspected not to be a ``genuine'' globular cluster but the remnant of
a dwarf galaxy that merged with the Milky Way in the past.  Because of
its proximity, this giant system is a cornerstone in our understanding
of the formation, chemical enrichment and dynamical evolution of
stellar systems.

\section{The central BSS population}
\label{obs}

 The results presented here are based on a high-resolution
sample obtained with the Advanced Camera for Surveys (ACS)
onboard the Hubble
Space Telescope (\HST) through $B$  and $R$  filters. The
observations are organized in a mosaic of $3\times3$ pointings
covering the $9'\times9'$ around the cluster center. Preliminary
results based on this data-base (dealing with the complex structure of
the SGB-TO region), have been already published \citep{F04a}. 
Here we focus our attention on the BSS population.

Figure 1 shows the zoomed color-magnitude diagram (CMD) in the BSS
region. As can be seen, there is a clear, well 
defined and populated BSS sequence,
cleanly separated from the MSTO stars. As
reference, the isochrones for 
 2 and 3 Gyr old populations \citep{C04} are also shown.
 We adopted a global metallicity ${\rm [M/H]
\simeq -1.5}$ (or $Z=0.0006$) \citep{F04a} for the metal-poor, dominant
population, a distance modulus $ (m-M)_0=13.70$ (i.e. $ \approx
5500$ pc, Bellazzini et al. 2004) and reddening
$E(B-V)=0.11$ \citep{L02}.  The mean interstellar
extinction coefficients listed in Table 2 by 
\cite{SM79} have been adopted.
  As can be seen,
the observed BSS sequence is nicely marked by the theoretical
sequences. In particular, the positions of the brightest BSS are
similar to MSTO stars  in 2--3 Gyr old
clusters, while the MSTO of the dominant metal-poor population, has
been fitted by  14--15 Gyr isochrone \citep{F04a}.
The MSTO masses in 2 and 3 Gyr old clusters are 1.4 and 1.2
$M_{\odot}$, respectively. This is in 
agreement with the recent finding 
of \cite{R04} who also concluded that the masses of BSS detected in
\ocen\ do not exceed 1.4 $M_{\odot}$. Indeed, the MSTO mass in \ocen\ itself
is estimated to be 0.74$M_{\odot}$, so mass transfer in an equal mass
binary should not produce a BSS more massive than 1.5$M_{\odot}$.

The detection of such a clean BSS sequence allows us to perform a
direct comparison with normal cluster stars.  In order
to be conservative and to avoid any possible contamination from
spurious blended objects (mainly due to MS stars), we 
selected only the brighter ($ 16<B<18.4$) and bluer ($B-R<0.7$)
portion of the BSS sequence.

Generally, in order to study the BSS radial distribution, both
horizontal branch (HB) and red giant branch (RGB) stars are used as
reference stellar populations.  In most clusters the HB is clearly
defined, and it has been used in previous papers \citep{F03} for
cluster to cluster comparisons. On the other hand, the RGB in the same
magnitude range of the BSS, including the lower RGB, is much more
populous than the HB, hence star counts are less affected by statistical
fluctuations. For this reason we used both populations as 
reference.   
Particular care has been devoted in selecting the 
RGB sample: {\it (i)} AGB stars are
 clearly separable from the RGB and are excluded from the reference sample; 
 {\it (ii)} we accurately selected RGB stars
belonging to the  dominant metal-poor branch. 
The possible contamination by  the more metal rich  RGB
stars is always negligible. 

Figure~\ref{rad} shows the cumulative radial
distributions of BSS and reference populations. 
A Kolomogorov-Smirnoff (KS) test indicates that the radial
distribution of the BSS has a $\sim
67$\% probability 
to be extracted from the same  parent distribution as
HB and RGB stars. {\it This is the very
first time that BSS have been found to share the same radial
distribution of normal stars of the parent cluster}.
Eventually, as can be seen from Figure 2, BSS in 
\ocen\ appear, if anything, even less concentrated than HB and RGB stars,
contrary to what observed in any other cluster.

To further investigate the radial distribution of BSS, we
computed the BSS relative frequency $F^{\rm BSS}_{\rm
HB}={{N_{\rm BSS}}/{N_{\rm HB}}}$ (where $N_{\rm BSS}$ and $N_{\rm
HB}$ are the number of BSS and HB stars, respectively) and studied its
behavior as a function of the distance from the cluster center.  In
doing this, we divided the sampled area into a set of concentric
annuli, and in each annulus we have counted the number of BSS and HB
stars.  The relative frequencies as a function of distance from the
cluster center are shown in Figure 3 ({\it upper panels}): the
distribution appears nicely constant over the entire extent of the
ACS sample\footnote{Errors in the relative 
frequency ($F^{\rm BSS}_{\rm
HB}$) of BSS  with 
respect to the HB, are given by 
$$\sigma_F = (F^2\sigma^2_{HB}+\sigma^2_{BSS})^{1/2}/N_{HB}$$
where $\sigma_HB$ and $\sigma_BSS$ are errors derived from 
the Poisson statistics.}.  Moreover, the
constancy of the BSS frequency at different distances from the cluster
center is independent of the choice of the reference population.  We
emphasize that this is the first time that a such behavior has been
found in a globular cluster.  In all previously surveyed
clusters the central regions show a significant overabundance of
BSS. Typically, the specific frequency drops by a factor of 4 or
more over a few core radii\footnote{As discussed in Section
3.1 the core radii of \ocen\ is $r_c=153''$.}.
 Note that
BSS have been found to be significantly more centrally concentrated
even in clusters with significantly lower central density than \ocen\ (see for
example NGC~288, Bellazzini et al. 2002).

\section{Sampling the entire cluster}

 In order to further investigate the peculiar distribution of  BSS
in this stellar system, we extended the analysis to the external
regions.  To do this we used the wide-field photometry previously
obtained by our group using the Wide Field Imager (WFI) on the
2.2m ESO-MPI telescope at the European Southern Observatory (ESO) on La
Silla.  The WFI is a mosaic of 8 CCD chips (each with a field of view
of $8'\times 16'$) giving a global field of view of $33'\times 34'$.
The images were obtained using $B$ and $I$  filters and have
been previously used \citep{P00}  to identify an
additional, previously unknown RGB (RGB-a, see also Lee et al. 1999).
Since the cluster center is slightly off-center in that data-set, we
also supplement the sampling of the external region with 
the $B,~V$ catalog by \cite{R04}\footnote{The total region sampled 
by Rey et al. (2004) covers 
 $40'\times 40'$.}.
The final catalog samples
the cluster population over (complete) concentric annuli up to a
distance of 20 arcmin from the cluster center.

All the catalogs were transformed to the same absolute astrometric
system by using the Space Telescope {\it Guide Star Catalog II}, and
then photometrically matched by using the stars in common. Note that
all three datasets have the $B$ passband in common.  To be
conservative and in order to prevent any possible incompleteness
due to poor spatial resolution, we
excluded the innermost region ($r<8'$) of the WFI sample and used the
Rey et al. catalog only in the most external region not sampled by our
WFI observations.

\subsection{The star density profile}

As a first step, we used 
the selected sample of RGB/SGB in the magnitude range
$16<B<18.4$ to obtain a new density profile for the cluster. The
analysis has been performed using almost 23,000 stars.  The surveyed
region has been divided in 27 concentric annuli.\footnote{The use of
circular annuli is a good approximation. 
Even though \ocen\ is the
most elliptical GGC, Pancino et al (2003) showed that
the axial ratio of
the entire RGB  distribution changes only a few percent
in the
central region ($r<5\arcmin$).
 In the outer region statistical
fluctuations overwhelm any errors caused by the ellipticity.} Each
annulus has been split in a number of subsectors (generally octants or
quadrants). Then the number of stars lying within each sub-sector has
been counted and averaged. Star density has been obtained by dividing
the average star number by the corresponding sub-sector area (in ${\rm
arcsec^2}$). The resulting profile can be nicely reproduced
by a King model  characterized by two
parameters: the core radius, $r_c$ and the concentration, $c$.  We
redetermined these parameters using our observations and the projected
star density from a standard isotropic single-mass King model
\citep{SP95}. The result is shown in Figure 4: we found 
$r_c=153\arcsec=2\farcm 57$ and $ c =1.31$ in good agreement with the
values ($c =1.24$ and $r_c=155\arcsec$) listed by \cite{TKD95}.%%22

\subsection{The BSS population over the entire cluster}

We selected the BSS following the
criteria adopted for the ACS sample. Only BSS in the magnitude range
$16<B<18.4$ were considered. We used isochrones \citep{C04} to convert
the red edge of the selection box from the $(B-R)$ color into $(B-I)$
and $(B-V)$ colors. We found that the adopted red edge in the ACS
sample ($ B-R=0.7$) corresponds to $(B-I)\approx 1.0$ and
$(B-V)\approx 0.42$.
 
Figure 5 shows the zoomed CMDs in the BSS region for the ACS and WFI 
samples. The selection boxes are shown in each
panel. The BSS population appears as a clear sequence
diagonally crossing each box. To be conservative, we selected 
only stars within the two dashed lines. This accounts
for the bulk of the BSS populations with only a few
objects being excluded from the selection.  Following these
criteria\footnote{Note that slightly changing the box
boundaries has negligible impact on the overall results.}  
we have identified 158 BSS in the HST-ACS and 155 in the
external sample ($r>8\arcmin$, with 117  found in the region sampled by
the WFI and 38 in the Rey et al. sample), for a total of
313 BSS.  This is the cleanest and largest BSS sample detected in the
cluster. However, the global BSS population in \ocen\ probably is
significantly larger ($>400$) since {\it (i)} we sampled  $\approx
70\% $ of the cluster light; {\it (ii)} we limit our selection to the
brightest and cleanest portion of the BSS sequence.
 
Of course each photometric sample has its own incompleteness, and for
this reason we analyzed them separately. According to
section 2, each BSS population is
referred to the RGB population of the corresponding {catalog
in the same magnitude range ($16<B<18.4$) of the selected
BSS}.  

  To compare the samples we used the doubly normalized ratio $R_{BSS}$
 which gives the fraction of BSS counted in concentric annuli at
 different distance from the cluster center with respect to the
 fraction of light sampled in each annulus. The sample presented here
 covers the cluster extent up to $20\arcmin$ from the center.  Hence,
 the entire surveyed area has been divided into six concentric annuli
 excluding only the region between $320\arcsec$ and $480\arcsec$.  The
 cluster light sampled in each annulus has been
 computed by integrating the King profile \citep{king} fitted to the
 observed cluster density profile shown in Figure 4.  The same ratio
 has been computed for the reference population.  As emphasized in
 \cite{F93}, for a stellar population which is distributed accordingly
 with the integrated cluster light, this ratio is 1. The result is
 shown in Figure 6, and as can be seen the normalized BSS ratio is
 nicely constant and fully consistent with the reference population.

In other clusters (M3: Ferraro et al. 1997, 47~Tuc: Ferraro et
al. 2004b, NGC~6752: Sabbi et al 2004, M55: Zagggia et al. 1997) not
only does the normalized BSS population peak at the center, but it
again rises in the cluster periphery. This effect probably arises
because primordial binaries whose orbits are confined to the cluster
exterior have low collision rate and thus remain in the exterior where
some eventually become BSS. BSS progenitors at intermediate radii
drift inward where they are either ``ionizated'' or driven to merge by
collisions. Dynamical simulations have demonstrated this phenomenon in
47~Tuc (Mappeli et al. 2004).

Since this result seems to be quite peculiar, a natural question
arises: Is there any other evidence that can support our findings?
First we note that there is no convincing evidence of equipartion
effects in the radial distribution of MS stars of \ocen\ \citep{A02}.
What about the radial distribution of other sub-populations
significantly more massive than normal cluster stars?  Interacting
binaries containing a compact object (like a neutron star or white
dwarf) in which mass transfer is occurring are expected to show X-ray
emission. These are among the most massive objects we can currently
find in an old stellar population. We used the recent XMM-Newton
observations \citep{G03} in order to check it. There are 42 X-ray
sources lying in the field-of-view covered by the ACS
observations. The radial distribution of these sources is compared
with that of BSS in the right panel of Figure 2. As can be seen the
radial distribution of the X-ray sources is the same or even less
centrally concentrated than the BSS.\footnote{Interpreting the
distribution of X-ray sources is complicated by the fact that some
could be background active galaxies).} A KS test shows that
BSS and X-ray sources have a $\sim 12\%$  probability of being extracted
from the same parent distribution. However due to the small sample of
detected X-ray sources, we can conservatively conclude that the
difference between the two distributions is not significant.  In
conclusion neither of the two most massive star populations 
in the cluster (BSS
and X-ray sources) shows any signature of radial segregation
with respect to the normal stars in the cluster.

Since relaxation is the major
dynamical process which differentiates stars according to their
mass \citep{MH97}, the observational facts presented here
represent the cleanest evidence found so far that the system is still
far from being completely relaxed even in the core region.

\section{Discussion}

The evidence presented here demonstrates that \ocen, the
largest  stellar system of the galactic halo, does not share the
same dynamical characteristics of GGCs. Indeed because of
its mass, previous hints and arguments have led, in the
past, to the
suspicion that \ocen\  is not completely relaxed  
\citep{Ma97}. Since
\ocen\ is one order of magnitude more massive than a typical halo GGC,
its relaxation time-scale is also expected to be significantly
larger.  Previous estimates \citep{D93} of the central relaxation time
for \ocen, give $ \log\,t_{\rm crt}=9.76$ i.e., $ t_{\rm crt} \approx
5.7 $ Gyr, adopting a distance of $d\approx 4900$ pc. This distance is
shorter than that currently adopted here ($d\approx 5500$ pc).  For
this reason we re-computed the relaxation time for this stellar
system, adopting the new distance and the structural
parameters obtained in Section 3.1. Under these assumptions, the
physical size of the core radius turns out to be $r_c\sim 4.1$ pc. 
By adopting the observed integrated
magnitude $ V_t=3.68$ \citep{H96}, the
resulting central relaxation time is $\log\, t_{\rm crt}=9.82$, i.e
$t_{\rm crt}\sim 6.6$ Gyr.  Although larger, this new determination of
the relaxation time is still a factor of two shorter than the cluster
age (12--14 Gyr), hence some segregation should be visible at least in
the central regions.  A good indication of the expected segregation
time-scale for an object of mass $m_*$ can be obtained by considering
the half-mass relaxation time-scale \citep{D04}. Using Equation (10)
of \cite{D04} we found that roughly half of the $m_*=1.0\, M_{\sun}$
BSS should have been sunk to the core after only 2 Gyr.

{\it Is the lack of segregation connected to the origin of this
peculiar stellar system? Can the history of a stellar system influence
its {\it internal} dynamical status? Is it due to rotation? }

According to Davies et al (2004) heavier stars tend to take longer
to sink in the core of more massive systems. If we accept the
hypothesis that \ocen\ is the relic of a dwarf galaxy partially
disrupted by the tidal field of the Milky Way, then this system was
significantly more massive in the past than what we observe today.
According to \cite{T04} the initial mass of \ocen\ could have been $M
\approx 10^8~M_{\odot}$ and the cluster should have remained more
massive than $10^7~M_{\odot}$ for a few Gyrs.  Equation (10) of Davies
et al (2004) suggests that the {\it sinking-time} is a factor 7--8
longer for a system $10^2$ times larger than the current \ocen.
In such a scenario, the stormy past of this stellar
system could have  extended the {\it
sinking-time} needed for the heaviest population to reach the
cluster center to a time larger than the cluster age.

Rotation can also play a role. Relaxation time is expected to be
longer for rotating systems, since angular momentum tends to keep
stars out of the core, counteracting mass segregation \citep{S01}.
There is evidence for large rotation of the \ocen\
system \citep{MM86,MMM97,vLLP02}.  Moreover, while the metal poor giants
(${\rm [Ca/H] < -1.2}$), belonging to the dominant cluster population
form a rotating system, the metal rich ones do not show any
significant rotation \citep{NFM97}. Such evidence suggests a
clear correlation between dynamical and chemical properties of
\ocen. Perhaps, the lack of equipartition observed could be related to
the origin of the multiple populations inside the cluster.

Although, we do not fully understand why \ocen\ is not
relaxed yet, we
still feel safe in concluding that stellar collisions have played a
minor role in generating exotic binary systems in \ocen.  We have
previously made detailed studies of nine clusters. Our major
conclusion is that the overall population of BSS in a typical GGC
turns out to be a complex conglomeration of collisional BSS and mass
exchanging binaries. Still, while several puzzles remain, some firm
facts are emerging. Among these are that, except for possibly NGC~288,
all clusters have some collisional BSS, and those BSS are strongly
centrally concentrated. This is hardly surprising since the collision
rate is higher at the center, and since the BSS are more massive than
the typical cluster star they will not migrate outwards. It is
possible that a ``kick'' produced by the collision ejects them from
the core (Sigurdsson et al. 1994, but see the discussion in Mapelli et
al. 2004).  However, Figure 6 shows that the number of BSS nicely
scales with the sampled luminosity in the central region as well as in
the outer regions. It would be most remarkable if some combination of
collision rate and kicks managed to produce the observed {\it flat}
BSS distribution. Applying Occam's razor, the simplest explanation is
that the BSS observed today in \ocen\ are the progeny of primordial
binaries whose radial distribution has not been yet significantly
altered by collisions.

Recently Piotto et al (2004) and Davies et al (2004) from
a large survey of BSS in 56 GGCs showed that the total
number of BSS is largely independent of the cluster
luminosity and collision rate, suggesting   that 
no single process produces BSS in GGCs
(accordingly with Baylin 1995, Fusi Pecci et al 1992),
hence both binary evolution and collisions both could be
active processes in producing BSS in different
environments. In particular, 
Davies et al. (2004)
developed a model for the production of BSS in GGCs. In the low mass
systems ($M_V > -8$) BSS arise mostly from mass exchange in primordial
binaries. In more massive systems collisions produce mergers of the
primordial binaries early in the cluster history. BSS resulting from
these mergers long ago evolved away. Once the primordial binaries were
used up, BSS produced via this channel disappeared. In the cores of the
most massive systems ($M_V < -9$) collisional BSS are produced (Fig.~6
of Davies et al.).

Detailed cluster-to-cluster comparison has shown that the scenario is
much more complex than that proposed by Davies et al (2004). The
dynamical history of each cluster apparently plays a role in
determining the origin and radial distribution BSS content. For
example, {\it (i)} clusters with comparable luminosity show vastly
different BSS population, see for example the case of M3/M13 ($M_V
\sim -8.7$ and $-8.5$ respectively, see Ferraro et al, 2003) and
NGC6752/M80 ($M_V \sim -7.7$ and $-7.9$ respectively, see Sabbi et al
2004 and Ferraro et al 1999) pairs; {\it (ii)} studies of the BSS
population over the entire cluster extension (M3, Ferraro et al 1997;
47 Tuc, Ferraro et al 2004b; NGC~6752, Sabbi et al 2004) have further
supported the complexity of the emerging scenario showing that the
central population of BSS is only a component of the entire BSS
population of each cluster. In addition to a strongly centrally
segregated population, these GGCs also have a population of external
BSS which we argue arise from primordial binaries. At intermediate
radii there is a paucity of BSS of either type, because the collision
rate is low and primordial binaries settled to the center and probably
became BSS long ago.
   
How does \ocen\ fit into the Davies et al. picture?  Following that
scenario, since \ocen\ is very luminous, its primordial binaries would
have been destroyed long ago, while
a population of several hundred collisional BSS should populate
the core of the cluster (see Figure 6 by Davies et al
2004). Hence we would expect to see a population of
collisional BSS in the core and possibly a population of surving
primordial binaries in the outer cluster. This would possibly produce
a bimodal distribution as observed in other clusters, or even a single
centrally-peaked distibution but in any case it cannot produce the
flat distribution shown in Fig.~6.\footnote{Unless one assumed
{\it(i)} that the collision rate has the same efficiency in producing
collisional BSS both in the center and in the cluster periphery or
{\it(ii)} that the efficiency of the collision rate in the center and
of the survival rate of primordial binaries in the outer region is the
same.}  From Fig.~1 of Piotto et al. with $M_V \sim -10$ we would
expect $F_{\rm BSS} \sim 0.1$. Curiously both $F_{\rm BSS}$ and the
number of BSS are close to what we observe. The curiosity arises
because we argue that the BSS are originated from primordial binaries
and the Davies et al scenario would suggest they were all collisional.

For sake of comparison, we can compare the expected collision rate for
\ocen\ (under the assumption that
 rotation or the cluster history has not made the estimate
 invalid)
to that expected in other massive GGCs such as 47
Tuc. According to Davies et al (2004) the collision rate is
$\Gamma_{coll} =5\times 10^{-15} \times (r_c \Sigma_0^{3})^{1/2}$ in
units of collisions per year (where $\Sigma_0$ is the central surface
brightness in units of $L_{\odot V} pc^{-2}$ (equivalent to
$\mu_V=26.41$) and $r_c$ is the core radius. By assuming
$r_c=0.42\,$pc (Mapelli et al 2004) and $r_c=4.1\,$pc for 47 Tuc and
\ocen\ , respectively, we find that the expected collision rate is
only 5--8 times larger in 47 Tuc as compared to \ocen\ (approssimately
50--60 collisions per Gyr in the core of 47~Tuc with respect to 6--11
in the core of \ocen\ ) depending on the adopted central surface
brightness values (Harris 1996 or Djorgovski 1993). In any case this
difference does not seem sufficient to explain the quite different BSS
content in the two clusters\footnote{Note that 47~Tuc harbor a
strongly centrally segregated population of BSS in the core (Paresce
et al 1991, Ferraro et al 2001) and a significant population of BSS in
the external region (Ferraro et al. 2004b)}.  This evidence further
supports the complexity of the BSS formation and evolution scenario
and suggests that the dynamical state of the cluster and its history
plays an important role in determining the BSS content.

Using the well populated BSS sequence in \ocen\ we have the
possibility to determine  the production rate of BSS
in a non-collisional system.  
From the upper panel of Figure 3 it is easy
to derive that the primordial-binary BSS specific frequency
$F^{BSS}_{HB}$ is 0.08.  Note that values  up to one order
of magnitude larger ($F^{BSS}_{HB}\approx1$) have been 
detected in the central region of GGCs \citep{F03}. 
An even more useful quantity for the study
of an unresolved stellar population is the ratio of the number of BSS
to the sampled luminosity in units of $10^4 L_{\odot}$, $S4_{\rm BSS}
=N_{\rm BSS}/ L_4$ \citep{F95}. $S4_{\rm BSS}$ varies
from cluster to cluster (ranging from 0.2 to 30) possibly
tracing the generation/destruction effect of stellar 
interactions in collisional
systems like the dense cores of GGCs. The bottom panel of Figure 3
shows the radial behavior of $S4_{BSS}$ within \ocen. The rate of
production is constant throughout the system despite the large variation
of stellar density.  If $S4^{\rm prim}_{\rm BSS}$ refers to primordial
binary BSS, our results suggest that $S4^{\rm prim}_{\rm BSS}\sim
2$. This means that in absence of collisions $\sim 2$ BSS can be found
for each $\sim 10^4 L_{\odot}$ of sampled light. This value can be
used to estimate the expected number of BSS generated by primordial
binaries for each fraction of sampled light in any stellar system.

Of course the specific frequency of primordial BSS is expected to
depend on the fraction of primordial binaries contained in the
considered stellar system. This fraction is still largely unknown, it
has been estimated only in a few traditional GGCs
(NGC~6752---Rubenstein \& Bailyn 1997, NGC~288---Bellazzini et
al. 2002, 47~Tuc---Albrow et al. 2001) and some lower density systems
(Pal~13---Clark et al. 2004, E3---Veronesi et al. 1996). In one of
those (47 Tuc) the determination has been obtained
 in the very dense central region,
where binaries are created, destroyed and hardened by stellar
collisions. It would be highly desirable to directly measure the
binary fraction in \ocen. Sadly the technique used by \cite{Be02} and
\cite{ruba1} it can hardly be applied to \ocen\ because its MS is broadened
by the abundance spread.

How does the BSS specific frequency in \ocen\ compare with
the field?  The  BSS  specific frequency ($F\sim 0.1$)
measured here turns out to be  40 times lower than
that ($F\sim 4$) found by Preston \& Sneden (2000)  in the
field. Indeed, such a large frequency  has never been
observed in any GGC.  One possible origin of this
discrepancy is that GGCs never harbored the sort of
primordial binaries that produce the long period BSS
binaries found in the field by Preston \& Sneden.  However,
it must be noted that their value is  only an indirect
estimated derived from a chain of assumptions based on
the   observed fraction ($60\%$) of the blue metal poor
field stars.  

Indeed the low BSS frequency compared to the field  and
the lack of segregation in \ocen\ are hard to reconcile.
Strong cluster rotation, for example, could depress the
frequency of collisional BSS but would probably not affect
the  primordial binary population of the cluster. Then, if
the data in the field are eventually confirmed, the observations
presented here strongly  point  at a substantial difference
between the binary population in the field and in the
cluster.

As final consideration, we note that 
there is now a growing interest in determining the age of distant
stellar systems from their integrated spectra. The main age
discriminants arise from the stars at the MSTO. Since BSS are both
hotter and more luminous than MSTO stars, a significant population of
BSS can make a system appearing younger. Because of the stellar density in
galaxies is low enough that collisional BSS are probably not
important, BSS in galaxies should have properties similar to those
detected in \ocen. Hence, this  BSS population can represent
 a suitable  template for
estimating the contribution of BSS to the global spectral energy
distribution of distant galaxies.

\acknowledgements

{We thank an anonymous referee for his/her suggestions
which significantly improved the presentation of this paper.}
We  warmly thank Paolo Montegriffo for assistance during
the astrometry procedure. The financial support of the 
  MIUR (Ministero
dell' Istruzione, dell' Universit\`a e della Ricerca) is
kindly acknowledged. RTR is partially supported
by STScI grant GO-8709 and NASA LTSA grant NAG 5-6403.

\clearpage 
\begin{figure}
\plotone{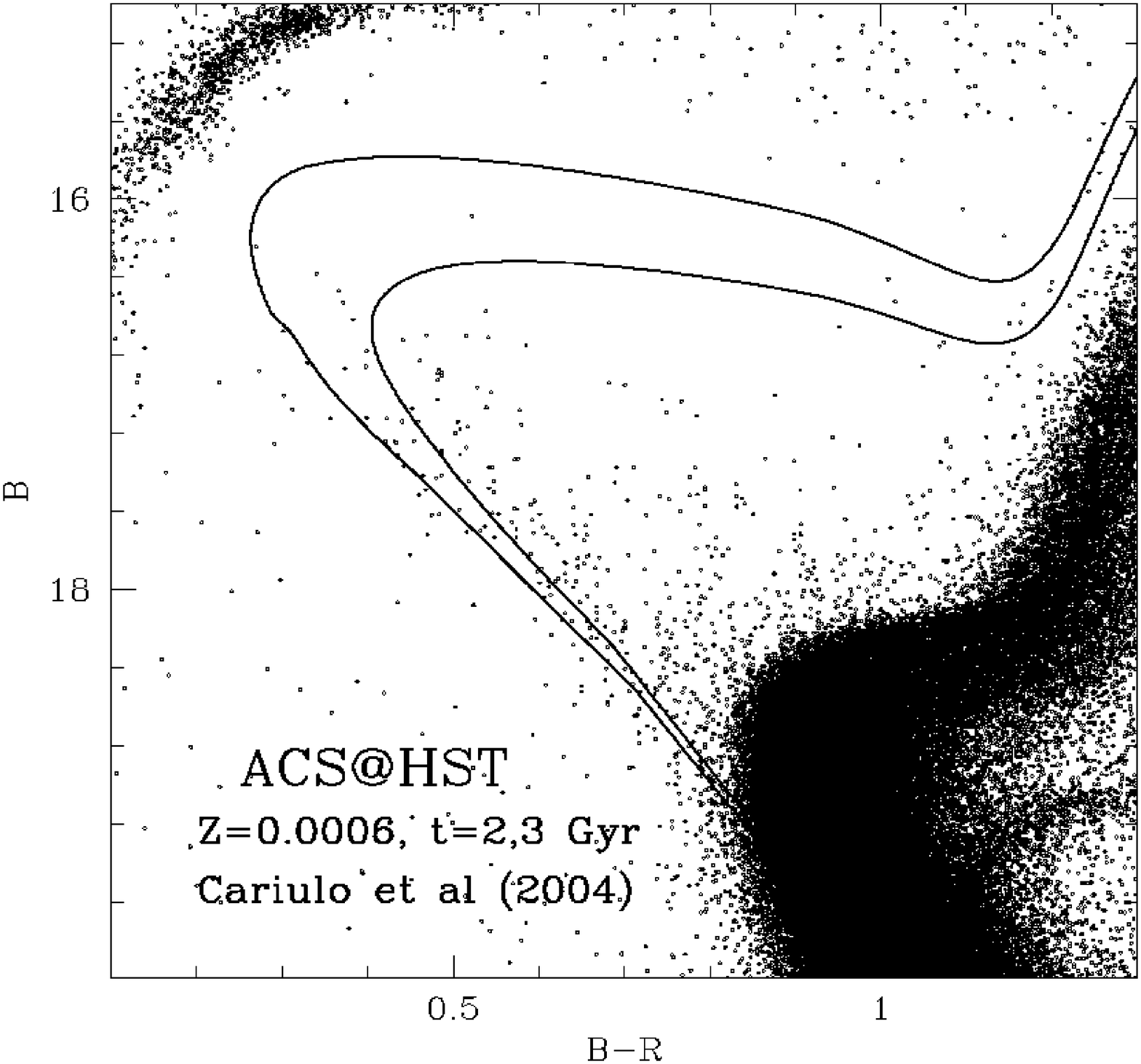} 
\caption{\label{iso} Zoomed CMD in the BSS
region for the ACS/HST sample. Two isochrones (by Cariulo et al. 2004) 
at Z=0.0006 and $t =2$ and $3$ Gyr are overplotted.}
\end{figure}

\clearpage

\begin{figure}
\plotone{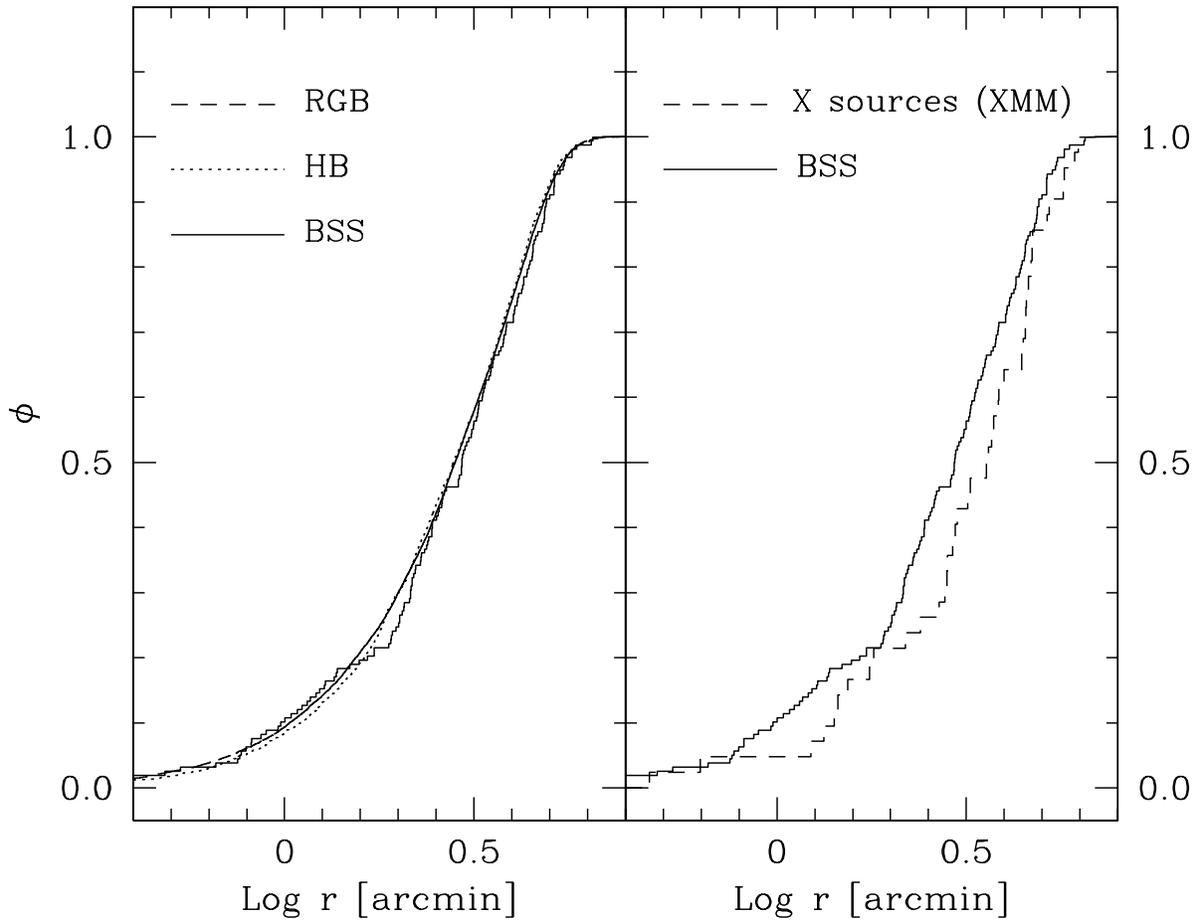}
\caption{\label{rad} {\it Left panel:} Cumulative radial distributions
for the BSS ({\it solid line}) in the ACS sample 
with respect to RGB stars ({\it
dashed line}) and HB stars ({\it dotted line}) as a function of their
projected distance ($r$) from the cluster center. {\it Right panel:}
Cumulative radial distributions for the cluster X-ray sources ({\it
  dashed line}) and BSS ({\it solid line}).}
\end{figure}

\clearpage

\begin{figure}
\plotone{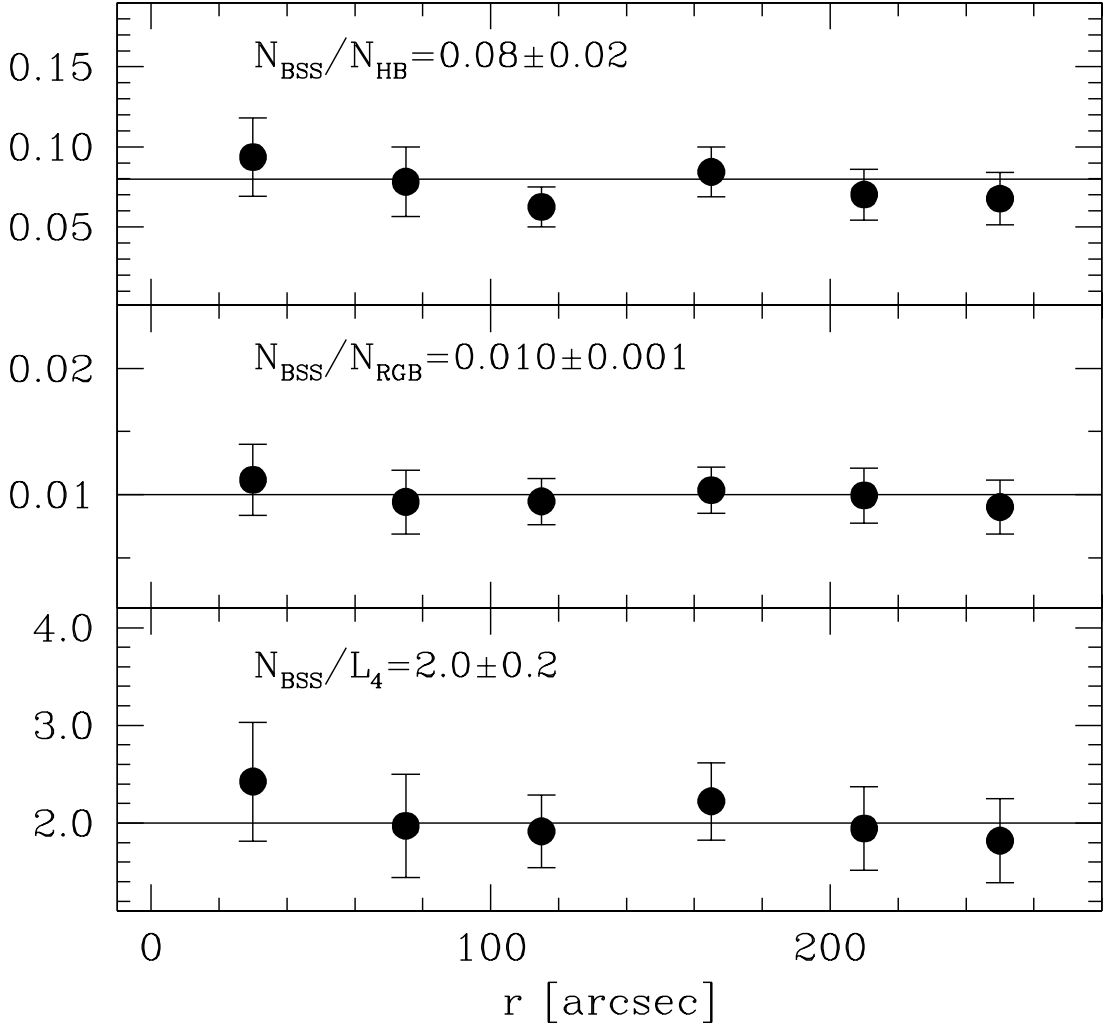}
\caption{\label{ratio} Specific frequency of BSS with respect to RGB,
HB stars, and the sampled luminosity $L_4$ in units of $10^4
L_{\odot}$ as a function of their projected distance ($r$) from the
cluster center. }
\end{figure}
 
\clearpage
\clearpage
\begin{figure}
\plotone{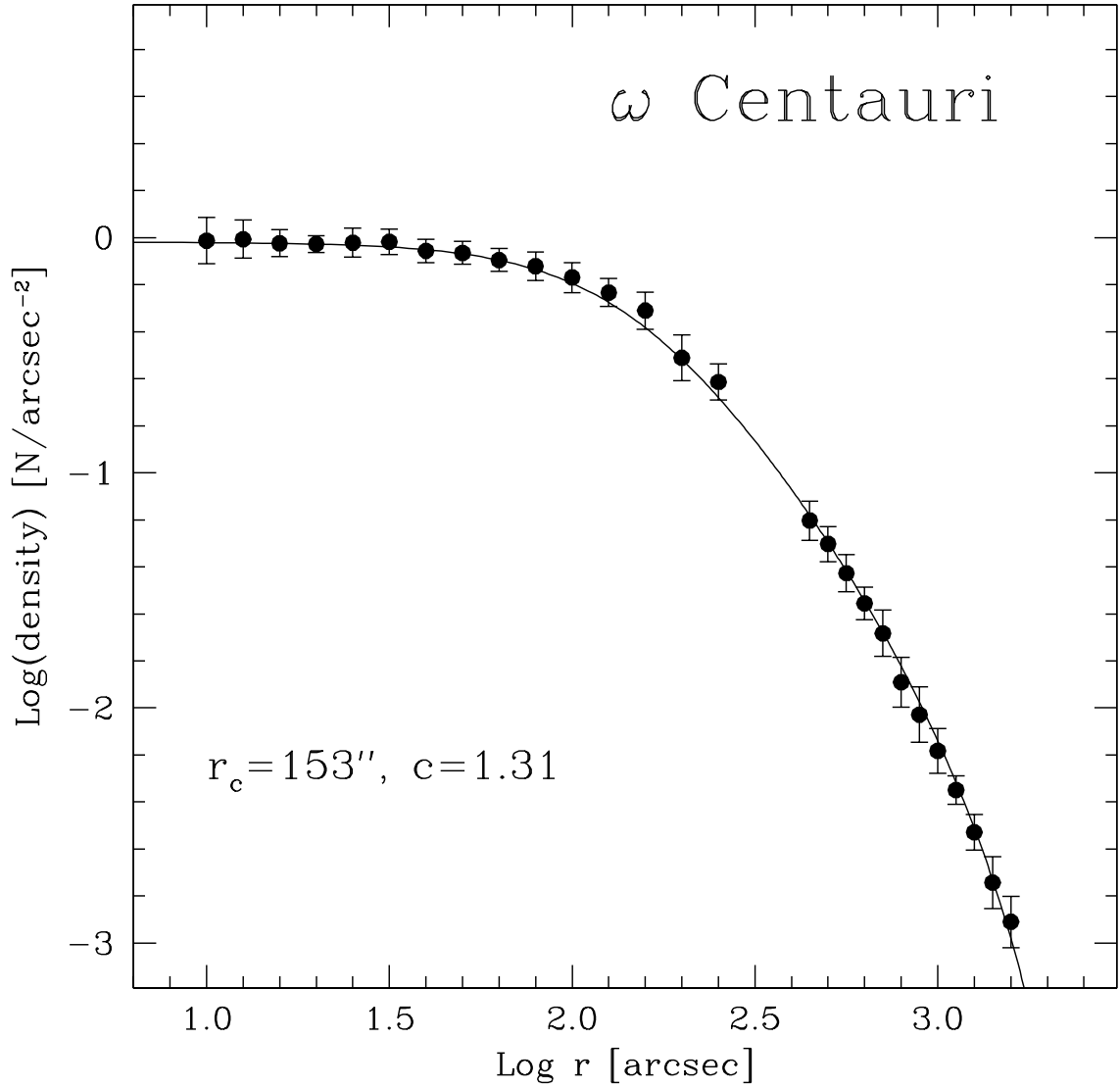} 
\caption{Observed radial density profile.
  The solid line is the best fit King model
($\rm r_c=153\arcsec$ and $\rm c=1.31$). }
\end{figure}

\begin{figure}
\plotone{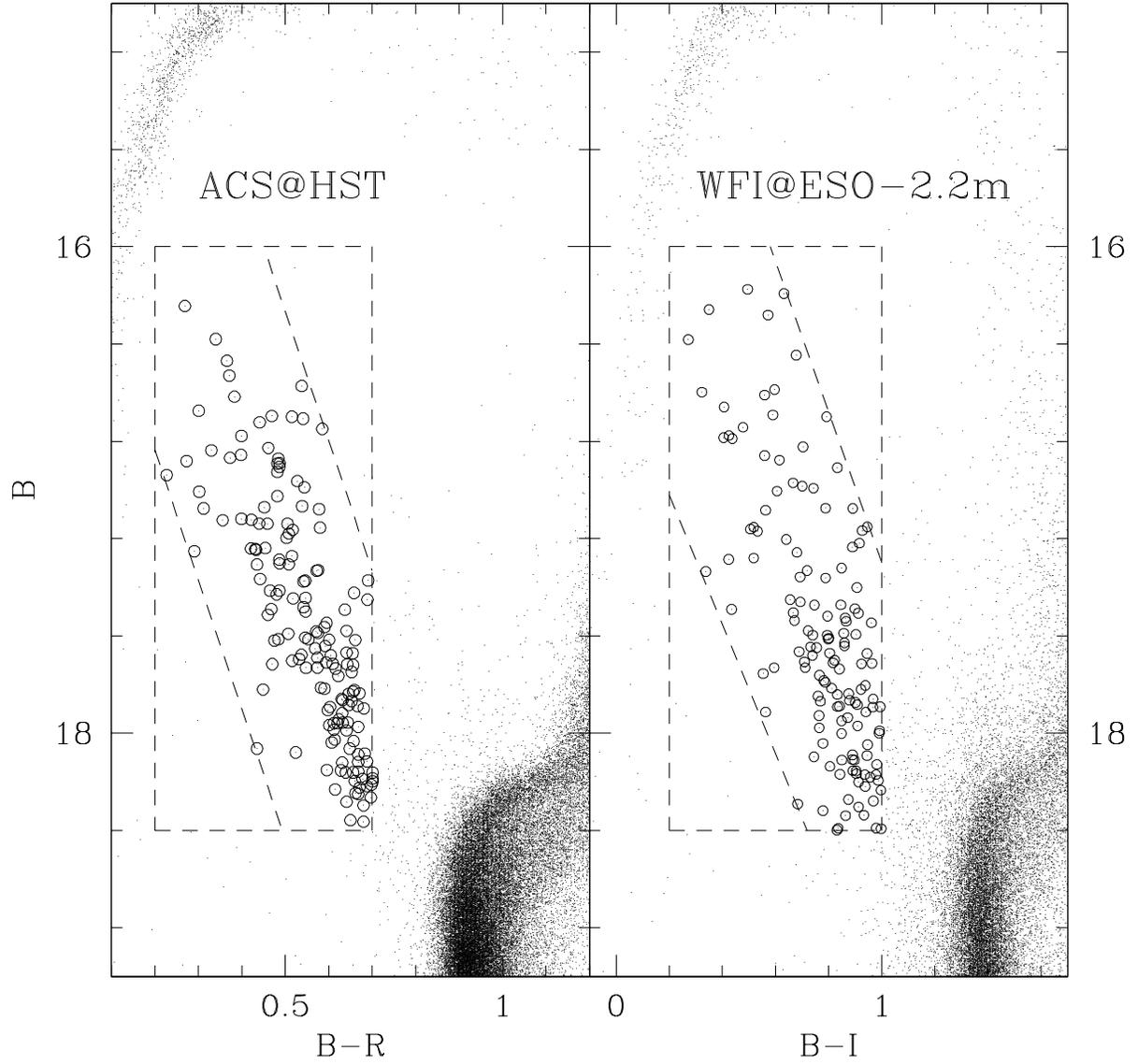} 
\caption{\label{hstwfi} 
BSS selection. Zoomed CMD in the BSS
region for the two samples:  
$B,~B-R$ CMD for the ACS/HST sample ({\it left panel}), 
and $B,~B-I$ CMD for the ground base WFI sample ({\it right
panel}). Only stars with $r>8'$ from the cluster center have
 been plotted in the WFI sample.
The selected BSS candidates  
are marked with large empty circles. }

\end{figure}

\clearpage
\begin{figure}
\plotone{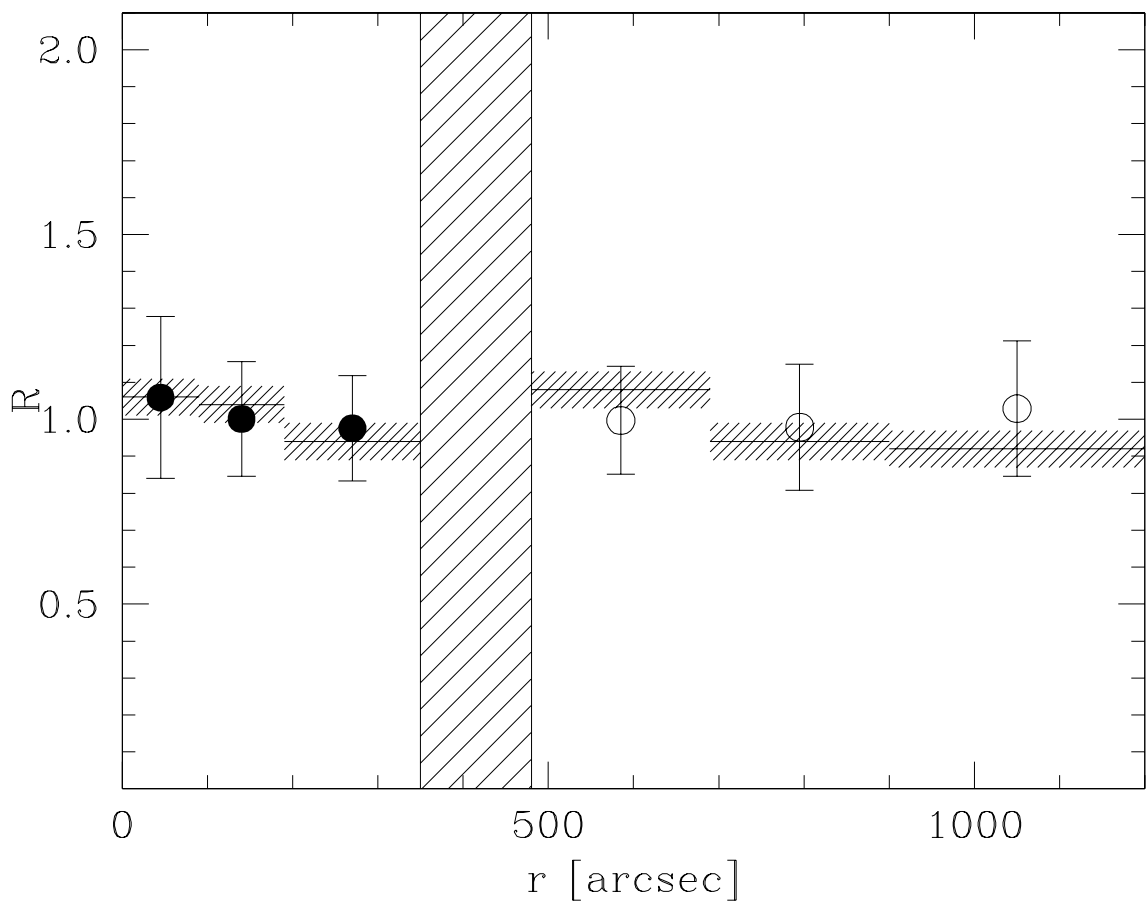}
\caption{\label{ratio1} The double-normalized relative 
frequency of the BSS $R_{BSS}$ (see text)  is plotted as a function
of the radial distance ($r$) from the cluster center. {\it Filled
circles} refer to the central region of the cluster
 surveyed with 
ACS/HST, {\it open circles} are combined 
(WFI and Rey et al.) ground-based
samples. The shaded area marks the cluster region 
 we excluded in order to avoid incompleteness
 problems. The horizontal  lines show  the relative frequency
 of the RGB stars used as reference population. 
  }

\end{figure}

\end{document}